\documentclass[11pt]{article}
\usepackage{latexsym,amsmath,amsfonts,graphicx}

\begin{document}
\begin{center}
{\LARGE\textbf{Consensus on Moving Neighborhood Model of Peterson Graph}}\footnote{This is the extended abstract prepped by HA.}\\
\bigskip
Hannah Arendt\footnote{ Institute of Mathematics, Technical
University of Liberec, Liberec, Czech Republic. email: haarendt@libimseti.cz}\\
and

Jorgensen Jost\footnote{Max Planck Institute for Mathematics in the
Sciences, Leipzig, Germany. email: jdjost@mis.mpg.de}\\

\end{center}

\begin{abstract}
In this paper, we study the consensus problem of multiple agents on
a kind of famous graph, Peterson graph. It is an undirected graph
with 10 vertices and 15 edges. Each agent randomly walks on this
graph and communicates with each other if and only if they coincide
on a node at the same time. We conduct numerical study on the
consensus problem in this framework and show that global consensus
can be achieved. \bigskip

\end{abstract}

\bigskip
\normalsize

\newpage
\section{Introduction}

In a consensus problem, a set of agents aim to make an agreement on
some quantities of interest via distributive decision making. The
information interactions are based upon local neighboring structure.
A consensus is said to be achieved if all agents in the system tend
to agree on the quantities of interest as time approaches infinity,
cf. survey papers \cite{3,5} and references therein. Especially, the
speed of synchronization is investigated in \cite{13}.

Ali Jadbabaie et al. \cite{1} studied a simple model of flocking
introduced by Vicsek et al. \cite{2} showing that all agents will
reach consensus as time goes on, provided the communication graph
switching deterministically over time is periodically jointly
connected. Some researchers have also treated random situations, see
e.g. \cite{6,8}. Recently, a new model called  moving neighborhood
network is introduced in \cite{10}. In this model, each agent
carries an oscillator and diffuses in the environment. The computer
simulation shows that synchronization is possible even when the
communication network is spatially disconnected in general at any
given time instant. Subsequently, several researchers have derived
analytical results on the moving neighborhood networks, see e.g.
\cite{18,9,11,4,12}.

The aim of this paper is to implement consensus on moving
neighborhood network modeled by the famous Peterson graph \cite{7}.
See Fig. 1. There are many interesting characteristics of Peterson
graph in mathematics. For example, it is traceable but not
Hamiltonian. That is, it has a Hamiltonian path but doesn't have a
Hamiltonian cycle. It is also the canonical example of a
hypohamiltonian graph. In this paper, we show that it is possible to
reach consensus on them by using moving neighborhood model.

\begin{figure}[hbt]
\centering
\includegraphics[80pt,653pt][214pt,771pt]{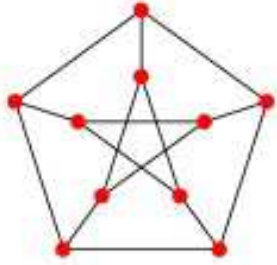}
\caption{An example of Peterson graph, which has 10 vertices and 15
edges.}
\end{figure}

\section{Preliminaries}

Let $G=(V,E,W)$ be a weighted graph with vertex set $V$. $E$ is a
set of pairs of elements of $V$ called edges. $W=(w_{ij})$ is the
weight matrix, in which $w_{ij}>0$ if $(i,j)\in E$, and $w_{ij}=0$
otherwise. Consider $n$ identical agents $\{v_1,v_2,\cdots,v_n\}$ as
random walkers on $G$, moving randomly to a neighbor of their
current location in $G$ at any given time. For each agent, the
random neighbor that is chosen is not affected by the agent's
previous trajectory. The $n$ random walk processes are independent
to each other. If $v_i$ and $v_j$ meet at the same node
simultaneously, then they can interact with each other by sending
information.

Let $X_i(t)\in\mathbb{R}$ be the state of agent $v_i$ at time t. We
use the following consensus protocol
\begin{equation}
X_i(t+1)=X_i(t)+\varepsilon\sum_{j\in
N_i(t)}b_{ij}(t)(X_j(t)-X_i(t))\label{1}
\end{equation}
where $\varepsilon>0$ and $N_i(t)$ is the index set of neighbors of
agent $v_i$ at time $t$. The factor $b_{ij}>0$ for $i\not=j$, and
$b_{ii}=0$ for $1\le i\le n$. Let $A(t)=(a_{ij}(t))$ be the
adjacency matrix of the moving neighborhood network, whose entries
are given by,
$$
a_{ij}(t)=\left\{
\begin{array}{cc}
b_{ij}(t),& (v_i,v_j)\in E(t)\\
0,& otherwise
\end{array}\right.
$$
for $1\le i,j\le n$. Suppose that $\triangle:=\max_{1\le i\le
n}(\sum_{j=1}^nb_{ij}(t))$, and we further assume
$\varepsilon\in(0,1/\triangle)$ for all $t$. We will show that the
states of all agents walking on a Peterson graph reach consensus as
time goes on.

\section{Numerical examples}

For Peterson graph represented in Fig. 1, we take the weight matrix
as the adjacency matrix. In addition, we take the $b_{ij}(t)$
randomly from a set of basic functions such as $e^t$, $\sin(t)$,
$\cos(t)$ and so on. In Fig. 2,3,4,5, we show that the consensus can
be achieved asymptotically.

\begin{figure}[htb]
\begin{center}
\scalebox{0.5}{\includegraphics{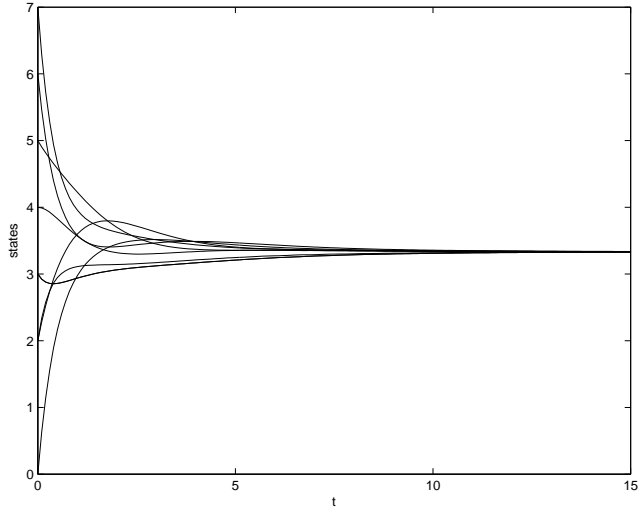}}\caption{The consensus
over moving neighborhood network modeled by a Peterson graph.}
\end{center}
\end{figure}

\begin{figure}[htb]
\begin{center}
\scalebox{0.5}{\includegraphics{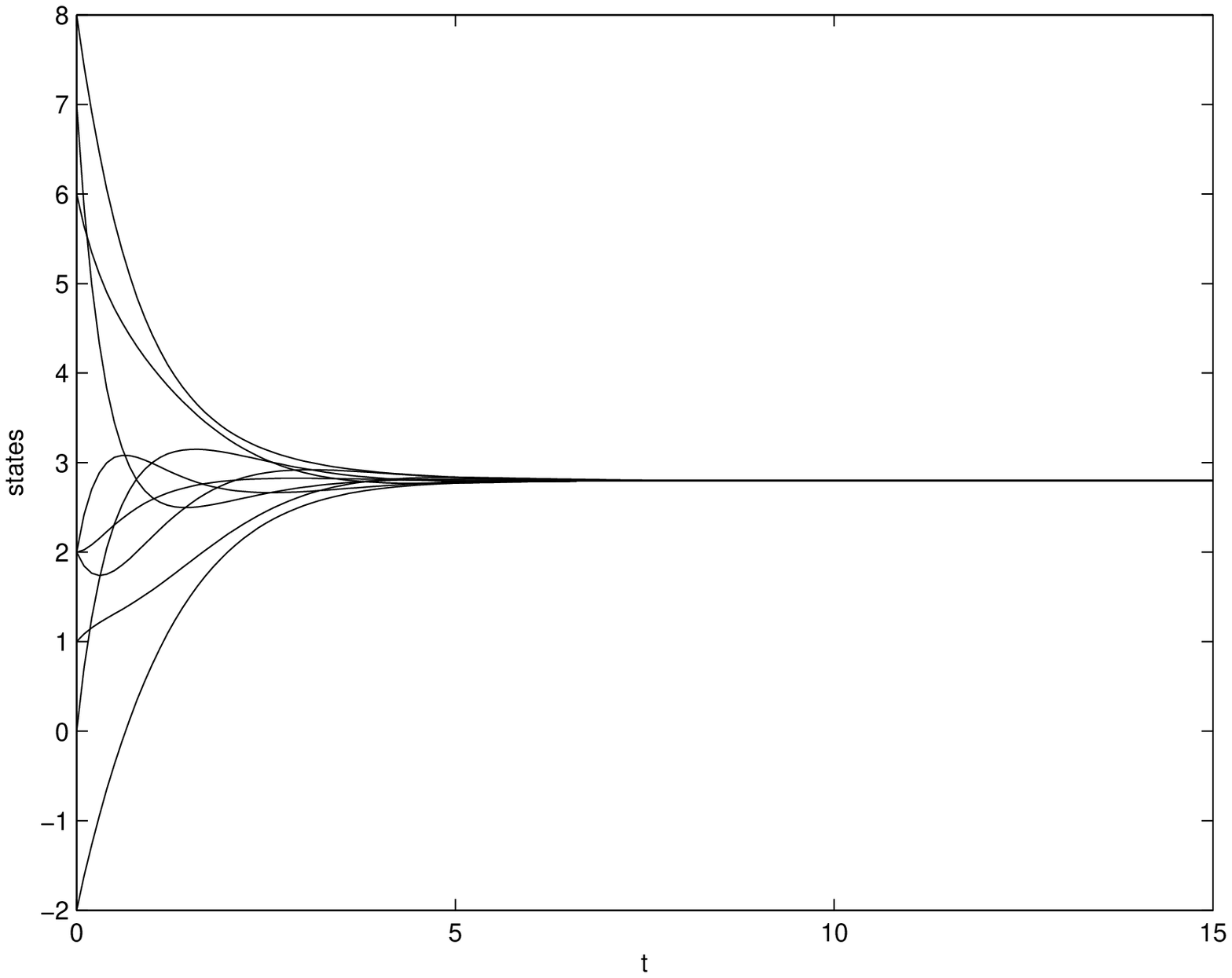}}\caption{The consensus
over moving neighborhood network modeled by a Peterson graph.}
\end{center}
\end{figure}

\begin{figure}[htb]
\begin{center}
\scalebox{0.5}{\includegraphics{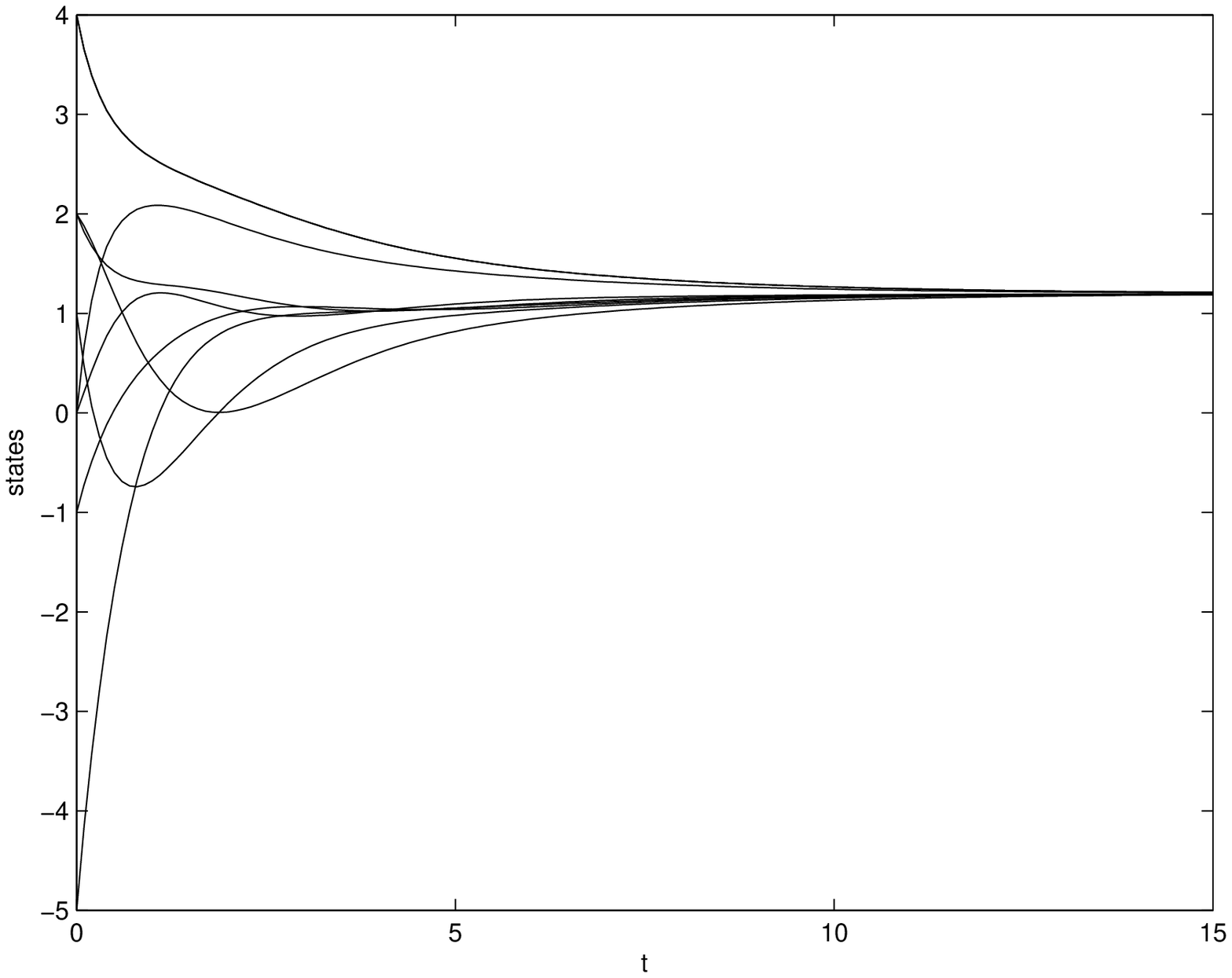}}\caption{The consensus
over moving neighborhood network modeled by a Peterson graph.}
\end{center}
\end{figure}

\begin{figure}[htb]
\begin{center}
\scalebox{0.5}{\includegraphics{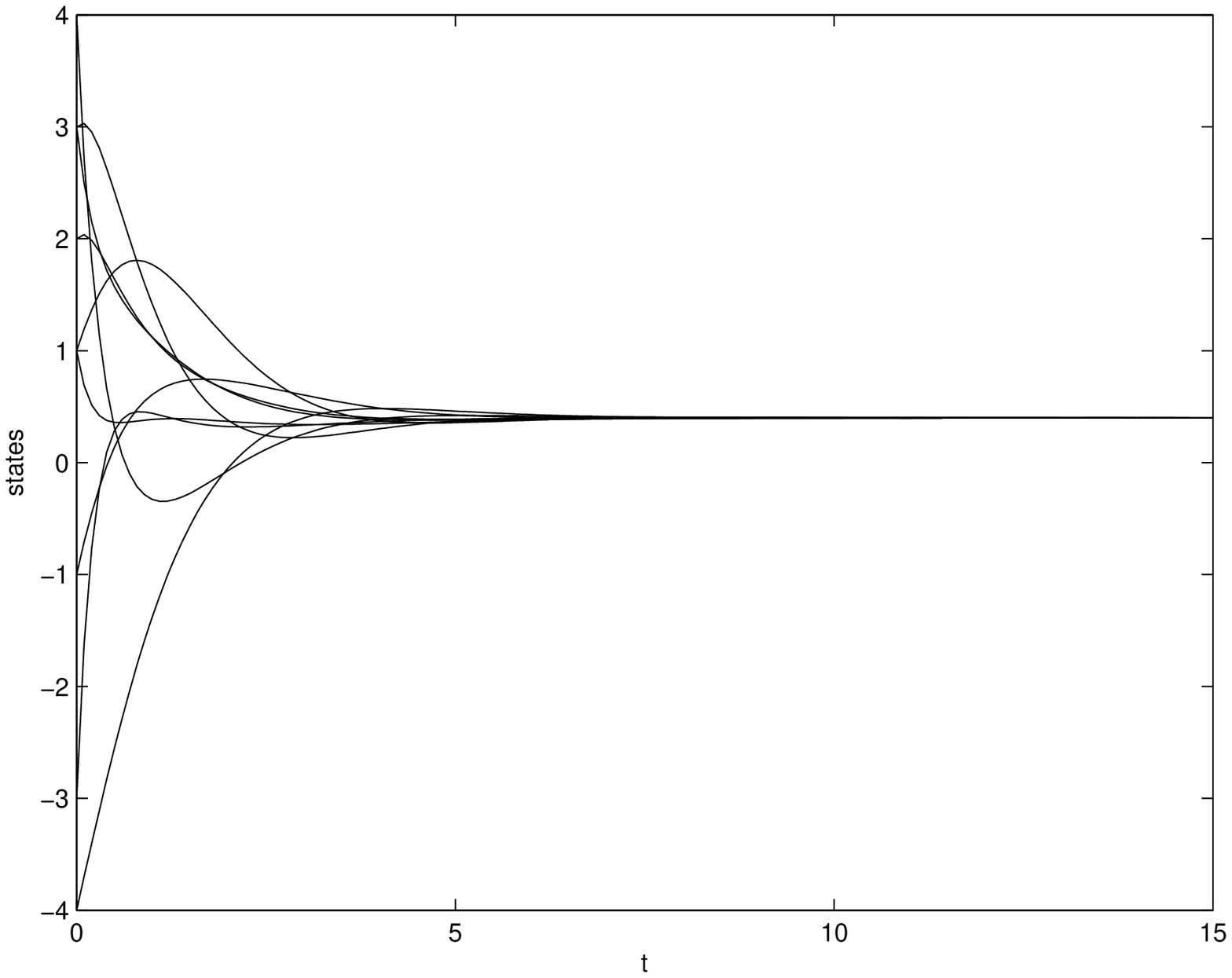}}\caption{The consensus
over moving neighborhood network modeled by a Peterson graph.}
\end{center}
\end{figure}

\end{document}